\documentclass[cameraready]{Interspeech}
\usepackage[margin=1in]{geometry}
\usepackage{amsmath, amssymb, amsfonts}
\usepackage{algorithm}
\usepackage{algpseudocode}
\usepackage{booktabs}
\usepackage{hyperref}
\usepackage{multirow}

\title{Pushing the Boundaries of Streaming Multi-Speaker ASR:\texorpdfstring{\\}{: } A Systematic Study of Architectural Trade-offs}

\renewcommand{\authorlist}{%
  \mbox{Taejin Park}, %
  \mbox{Ivan Medennikov}, %
  \mbox{Kunal Dhawan}, %
  \mbox{Weiqing Wang}, %
  \mbox{Jagadeesh Balam},
  \mbox{Boris Ginsburg}%
}
\address{
     NVIDIA, Santa Clara, CA, United States 
}

\email{taejinp@nvidia.com, kdhawan@nvidia.com, imedennikov@nvidia.com, weiqingw@nvidia.com, jbalam@nvidia.com, bginsburg@nvidia.com}


\usepackage{comment}

\begin{document}

\maketitle
\begin{abstract}
Streaming multi-speaker ASR is a challenging task that must balance accuracy, latency, and efficiency while handling overlapping speech and maintaining coherent long-context modeling over extended conversations in an online fashion.~We present a unified framework that categorizes streaming multi-speaker ASR into four architectural strategies based on how diarization and ASR are integrated. 
Using a shared pair of open-source streaming ASR and diarization models as a common foundation, we derive four multi-speaker ASR systems that differ in whether they employ multiple model instances, fine-tuning, or both. We evaluate these systems across multi-speaker accuracy, single-speaker accuracy degradation, memory footprint, and training complexity. Through this systematic architectural analysis, we clarify the design space for streaming multi-speaker ASR and provide practical guidance for selecting the most suitable approach under diverse deployment constraints.
\end{abstract}

\section{Introduction}
\label{sec:intro}

The demand for multi-speaker (or multi-talker) automatic speech recognition (ASR) has been rising dramatically, driven by applications such as voice agents powered by duplex models and speech-to-speech (S2S) systems~\cite{arora2025landscape}. In particular, interactive voice agent systems typically require low-latency streaming ASR, or a speech encoder integrated with an S2S model, in order to enable responsive conversational flow and deliver a seamless user experience.

While numerous studies have investigated multi-speaker ASR, the majority of prior work has focused on offline settings. Early systems~\cite{chang2019end, chang2019mimo} employed multiple encoders or attention heads to separate and recognize overlapping speech. More recent advances introduced end-to-end multi-speaker ASR based on Serialized Output Training (SOT)~\cite{kanda2020serialized}, which reformulates overlapped speech into a serialized sequence, allowing the multi-head self-attention mechanism~\cite{vaswani2017attention} to more effectively align acoustic streams with token sequences. Building upon this foundation, subsequent studies proposed several enhanced variants of the SOT framework~\cite{kanda2020joint,sotdom2024,fan2024sa}. More recently, token-level SOT~\cite{kanda22b_interspeech} has been developed, enabling multi-talker ASR in streaming scenarios.

In addition, the integration of speaker diarization models has led to highly modular architectures. A common approach is to employ a speaker diarization system that identifies and separates each speaker’s speech stream. The resulting segments are then processed by a single-speaker ASR model for transcription~\cite{medennikov2020stc}, where input features are masked to ensure that each ASR instance transcribes only one speaker at a time. Furthermore, approaches such as guided source separation (GSS)~\cite{boeddeker2018front} are often applied on top of diarization systems to estimate spectral masks, thereby improving the quality of the separation process.

Although these systems have demonstrated strong performance in challenge evaluations and under constrained experimental conditions, a comprehensive study quantifying the trade-offs between different architectural choices has not been thoroughly explored in the multi-speaker ASR community. When deploying streaming multi-speaker ASR systems in real-world scenarios, users encounter a wide range of practical constraints, including limited availability of training or fine-tuning data, restricted access to model parameters, and memory limitations. 
We identify the need for a systematic investigation of such trade-offs to enable researchers and practitioners to better tailor multi-speaker ASR systems to their specific requirements.

\begin{figure}[t]
  \centering
  \includegraphics[width=0.98\linewidth]{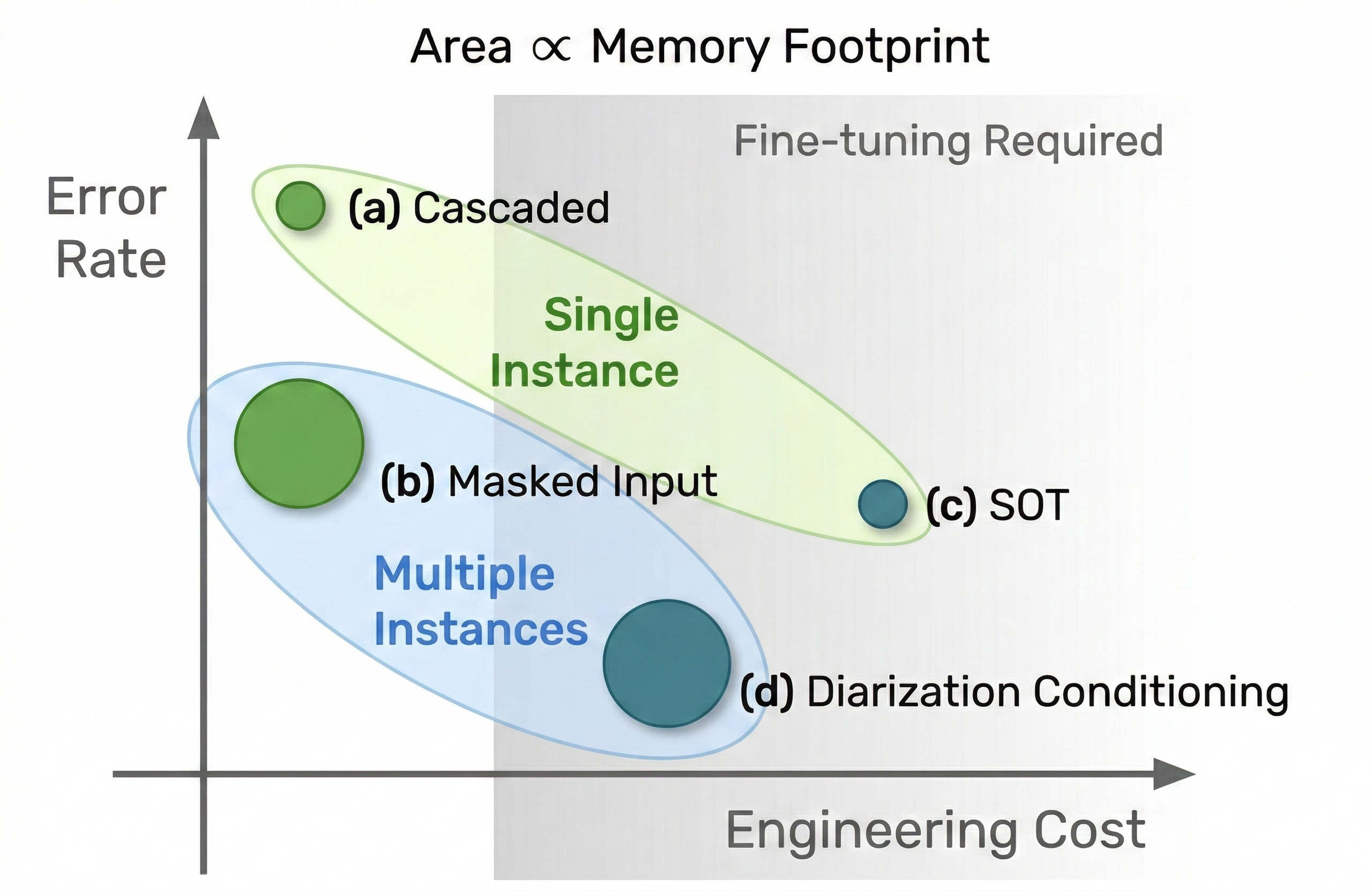}
  \vspace{-2ex}
  \caption{Trade-offs: Error Rate vs. Engineering Cost and Memory Footprint.}
  \label{fig:trade_off_plot}
  \vspace{-5ex}
\end{figure}

The contributions of this work are as follows:
\begin{itemize}
    \item We present a unified framework that systematically categorizes streaming multi-speaker ASR into four architectural paradigms, all derived from the same pair of open-source streaming ASR and diarization models, enabling the first controlled comparison across these design choices in the multi-speaker ASR community.
    \item We propose a transcription-to-token alignment algorithm that resolves the label-permutation ambiguity between diarization timestamps and SOT~\cite{kanda22b_interspeech} transcripts, removing a key obstacle to scalable SOT-style streaming multi-speaker ASR training data preparation.
    \item Through systematic evaluation on real-world conversational benchmarks---where, unlike prior work~\cite{kanda2022streaming} that predominantly reports on simulated mixtures such as LibriCSS~\cite{chen2020continuous} or operates in offline settings, we operate under strict streaming constraints on naturally occurring multi-speaker recordings---we quantify the trade-offs across multi-speaker word error rate, single-speaker accuracy degradation, memory footprint, and training complexity, and demonstrate that our best configuration achieves competitive accuracy while our alignment algorithm opens a viable path for scaling end-to-end SOT approaches.
\end{itemize}

\begin{figure}[t!]
  \centering
  \includegraphics[width=0.95\linewidth]{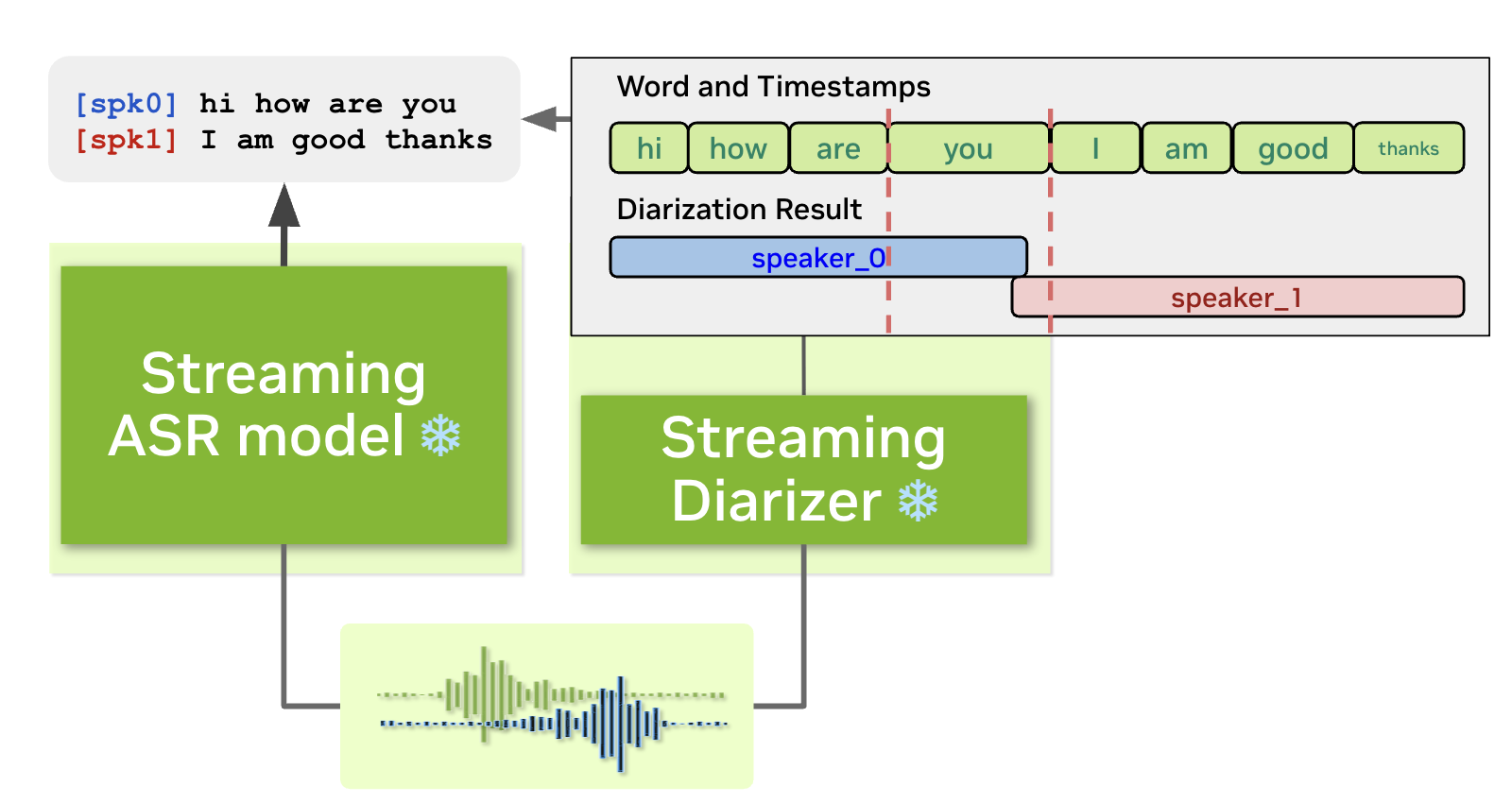}
  \vspace{-2ex}
  \caption{Cascaded approach: The two ASR and diarization systems do not have any interaction.}
  \label{fig:cascaded}
  \vspace{-5ex}
\end{figure}

\section{Architectural Trade-offs}

We categorize streaming multi-talker ASR systems into four representative architectural paradigms, distinguished by their use of multiple model instances and retraining requirements. We use the following open-source models: for the pretrained streaming ASR model, we use Nemotron Speech streaming ASR~\cite{nvidia_nemotron_speech_streaming_2025, noroozi2024stateful}, which is based on the Recurrent Neural Network Transducer (RNN-T) and FastConformer~\cite{rekesh2023fast} encoder; for the streaming diarization model, we use streaming Sortformer v2.1~\cite{nvidia_streaming_sortformer_2025, medennikov25_interspeech}.

\subsection{Cascaded ASR and Diarization}

Fig.~\ref{fig:cascaded} depicts the cascaded architecture, where diarization and ASR operate independently. Speaker attribution is achieved by mapping ASR word timestamps to diarization logits, assigning the speaker with the highest logit value to each word.
\begin{itemize}
    \item \textbf{Pros:}~Minimal engineering overhead and no retraining of either diarization or ASR models, making it practical when only API-level access to the recognizer is available. Strong modularity and compatibility with existing production systems, especially those built around standard streaming ASR APIs.
    \item \textbf{Cons:}~Limited ability to transcribe overlapping speech, since segments are handled independently and typically assume single-speaker dominance.~In addition, inaccurate word timestamps can cause boundary misalignment, leading to speaker assignment errors.
\end{itemize}

\subsection{Diarization Masked Input Architecture}

\begin{figure}[t!]
  \centering
  \includegraphics[width=0.9\linewidth]{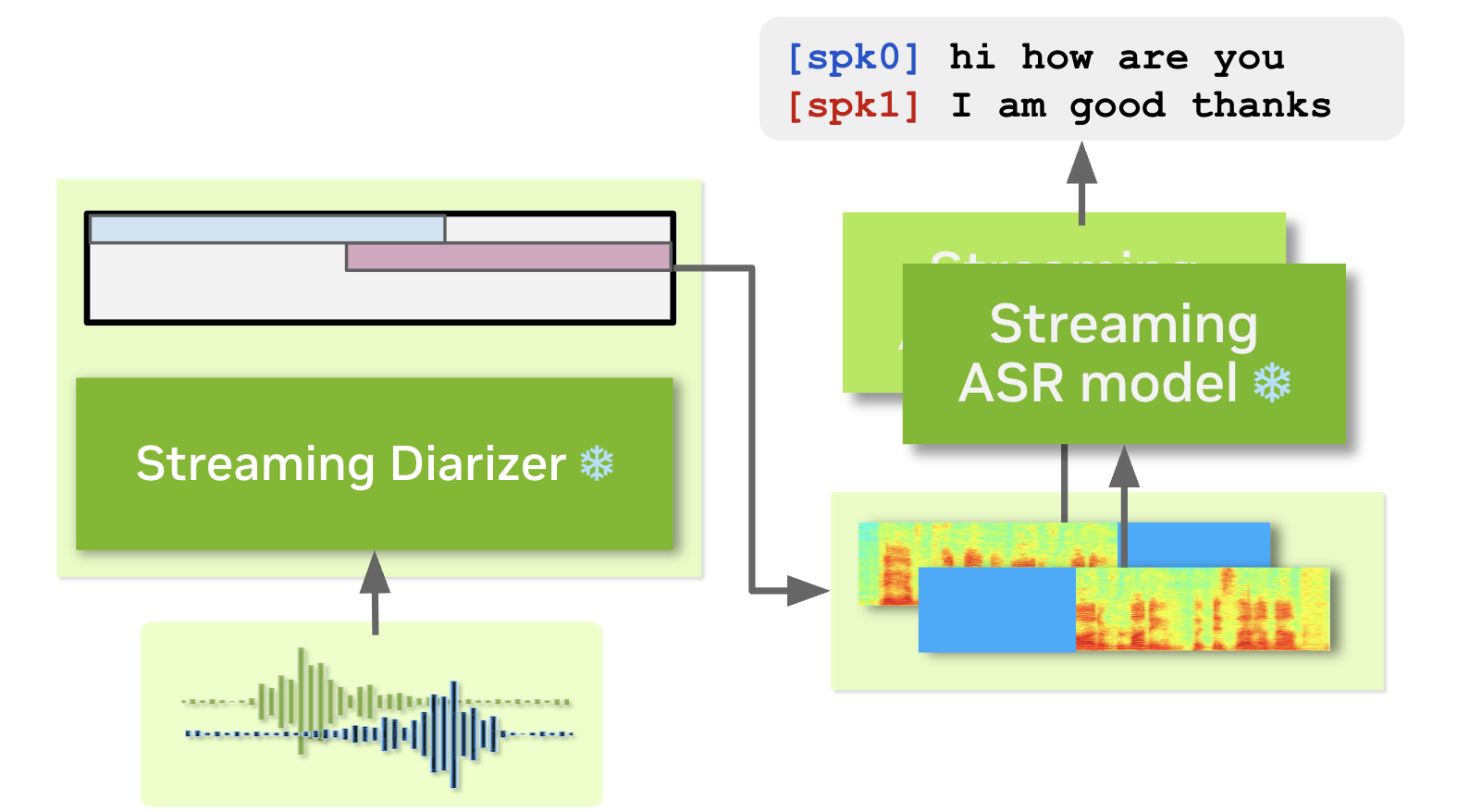}
  \vspace{-2ex}
  \caption{Masked Input Approach: Diarization information is provided at the audio feature input level.}
  \label{fig:masked_input}
  \vspace{-5ex}
\end{figure}

Fig.~\ref{fig:masked_input} shows the masked input architecture, where speaker-dependent masks derived from diarization outputs are applied to the acoustic features to enable parallel, multi-instance decoding for each speaker.

\begin{itemize}
    \item \textbf{Pros:}~Improved transcription of overlapping speech via masked feature streams that partially separate concurrent speakers. Reduced boundary misalignment since the method does not rely directly on ASR-generated timestamps for segmentation. No retraining is required, enabling the reuse of single-speaker ASR models and making the approach attractive for data-sparse languages while improving overlapping-speech recognition.
    \item \textbf{Cons:}~Overlap recognition can still lag behind systems explicitly fine-tuned for harsh multi-speaker conditions. The parallel, multi-instance setup may increase computational and memory costs as the number of speakers grows.
\end{itemize}

\subsection{Serialized Output Training (SOT)}

\begin{figure}[t!]
  \centering
  \includegraphics[width=0.85\linewidth]{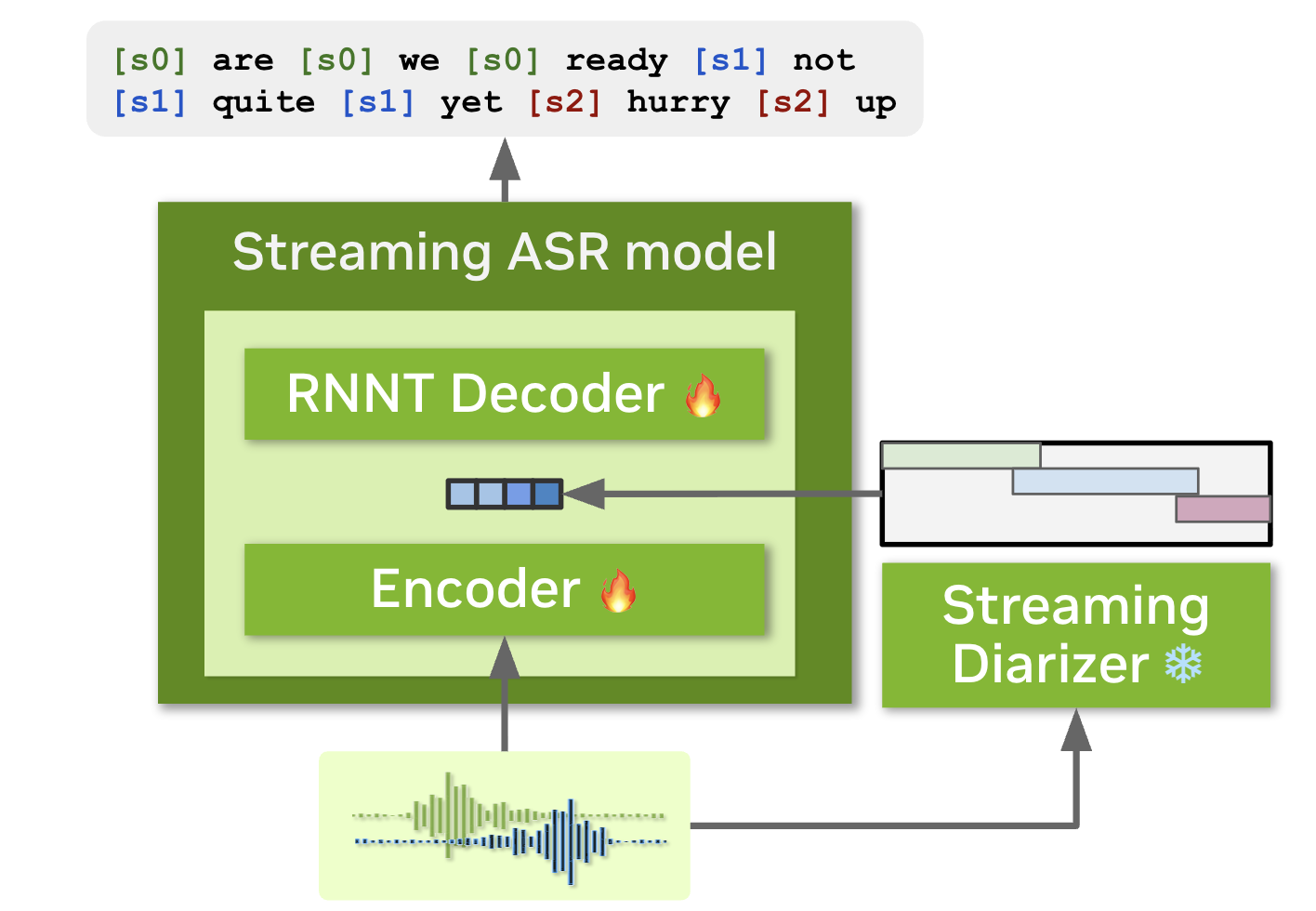}
  \vspace{-2ex}
  \caption{Serialized Output Training Approach: A single ASR instance and a streaming diarization model support long-term speaker cache management.}
  \label{fig:sot}
  \vspace{-3ex}
\end{figure}

The Serialized Output Training (SOT)~\cite{kanda2020serialized} approach paved the way for a fully end-to-end paradigm in which multi-speaker recognition leverages the attention mechanism to flexibly learn the alignments between acoustic and text sequences.
\begin{itemize}
    \item \textbf{Pros:} By serializing the multi-speaker transcriptions, SOT approaches show competitive performance while relying on only a single instance.
    \item \textbf{Cons:} Despite the success of SOT~\cite{kanda2020serialized} and t-SOT~\cite{kanda2022streaming}, there have been only a limited number of SOT-based streaming end-to-end ASR architectures~\cite{kanda2022streaming, subramanian25b_interspeech}, and these approaches have not demonstrated performance on real-world long-form audio streams. The primary reason is that end-to-end ASR training limits the ability to handle sequences longer than the training data (\textit{e.g.}, 12 s audio segments in LibriCSS~\cite{chen2020continuous}), making it challenging to generalize. In other words, \textit{the Train-Short Infer-Long} problem exists and is difficult to overcome with simple t-SOT and RNNT-based approaches.
\end{itemize}
To address long-form audio constraints, we propose a \textit{streaming word-level (WL)-SOT model} that integrates a speaker cache from a streaming diarizer. By grounding speaker prediction in an \textit{Arrival-Order Speaker Cache (AOSC)}~\cite{medennikov25_interspeech}, the model can maintain speaker supervision over indefinite durations. As illustrated in Fig.~\ref{fig:sot}, diarization predictions are embedded into the ASR encoder state via the speaker kernel mechanism introduced in~\cite{parksortformer}. However, training on short segments (10--60~s) extracted from long conversations presents a significant challenge: speech from previous turns often overlaps or spills over into the target segment. This results in a misalignment between speaker tokens and diarization timestamps (typically in Rich Transcription Time Marked, or RTTM, format), thereby hindering the model's ability to generalize from short training clips to long-form inference.

To address this, we introduce the Permutation-Invariant Dynamic Time Warping (PI-DTW) algorithm, which uses dynamic programming to align speakers and words simultaneously. Because DTW can struggle when word counts are imbalanced between speakers, we also incorporate a speaker frequency cost function. These two functions work together: DTW excels when speaking times are similar, while the frequency cost provides a clear distinction when one speaker dominates the conversation.

\begin{figure}[t!]
  \centering
  \includegraphics[width=0.8\linewidth]{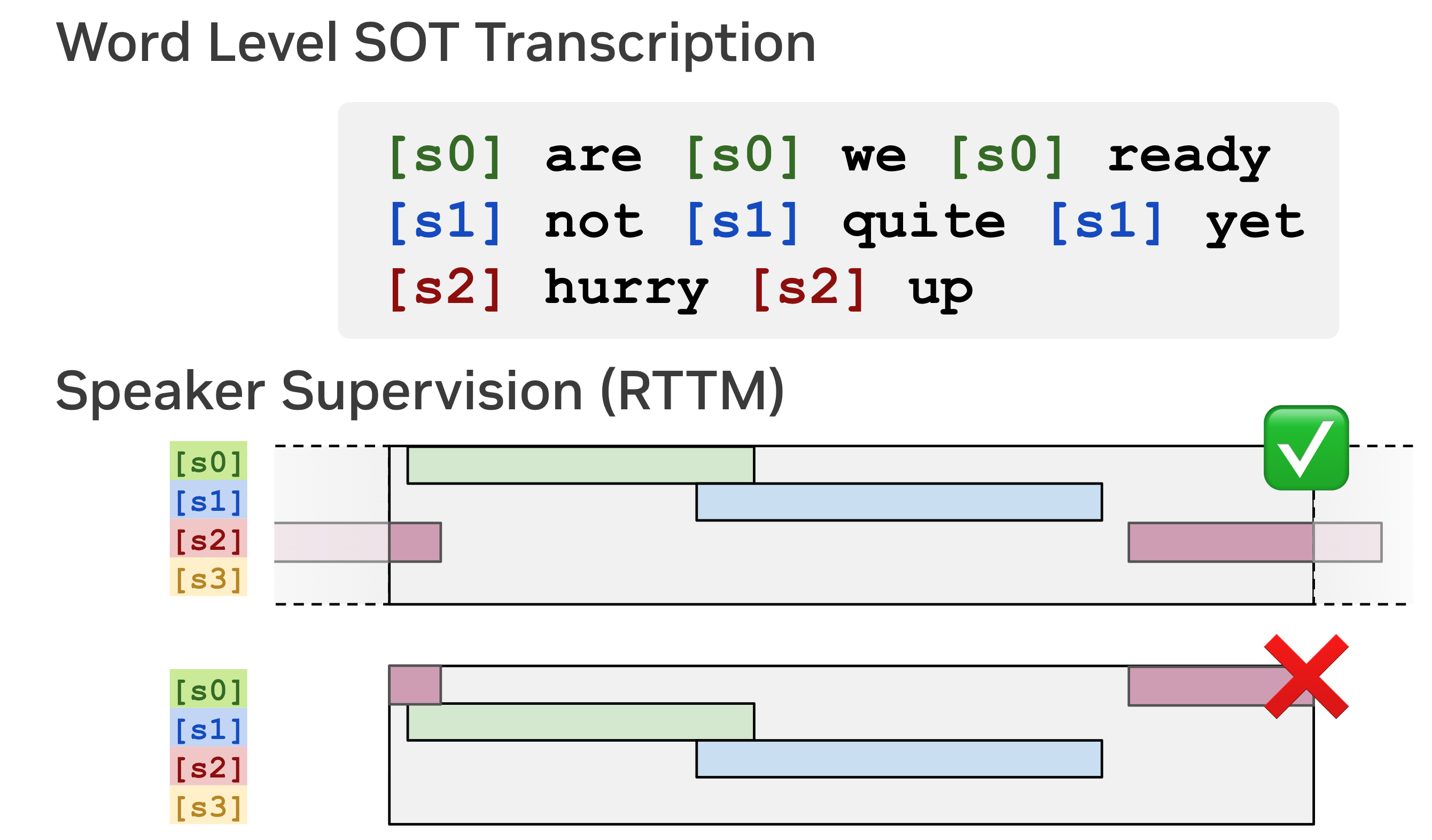}
  \vspace{-0.5ex}
  \caption{Example of speaker segment spill-over and transcription-RTTM mismatch. Applying arrival-time ordering can cause label mismatches.}
  \label{fig:sot_rttm}
  \vspace{-4ex}
\end{figure}

\subsubsection{PI-DTW Preliminaries: Notation and Local Cost}
Let $N$ be the number of speakers, $T$ the number of frames, and $K$ the number of words. $\mathbf{A}\in\{0,1\}^{T\times N}$ is the speaker activity matrix from RTTM files with $A_{t,j}=1$ if and only if speaker $j$ is active at frame $t$. $\mathbf{s}=(s_k)_{k=1}^K$, $s_k\in\{0,\ldots,N{-}1\}$ is the (per-word) speaker index sequence, and $\mathbf{w}=(w_k)_{k=1}^K$ are the words. $\pi:\{0,\ldots,N{-}1\}\to\{0,\ldots,N{-}1\}$ maps text speaker $i$ to RTTM column $\pi(i)$; define $\tilde{A}^{\pi}_{t,i}=A_{t,\pi(i)}$.

%% ---------------------------------------------------------------------------
% \subsubsection{RTTM-SOT Match Local Cost Function}

The local cost at grid cell $(k, t)$ measures the mismatch between the text speaker
one-hot vector and the permuted activity, normalized by the total activity at frame~$t$
to account for overlapping speech:
\begin{equation}
\vspace{-0.5ex}
c^{\pi}(k,t)
=1-\frac{\tilde{A}^{\pi}_{t,\pi(s_k)}}
{\max (1,\sum_{j=0}^{N-1} \tilde{A}^{\pi}_{t,j} )}.
\vspace{-1ex}
\end{equation}

The denominator $\sum_j A_{t,j}$ counts the number of simultaneously active speakers
at frame~$t$.  When exactly one speaker is active and it matches ($A_{t,\pi(s_k)}=1$),
the cost is~0.  When two speakers overlap and the text speaker is one of them, the denominator is 2, so the cost is $1 - \tfrac{1}{2} = 0.5$.  When the text speaker is inactive, the cost is~1.

%% ---------------------------------------------------------------------------
% \subsubsection{Speaker Token Count Weighting}

We apply inverse-frequency weighting so that each speaker contributes equally to the
total cost. Let $n_i$ denote the number of words attributed to speaker~$i$. The weight for word~$k$ (whose speaker is $s_k$) is $\omega_k = K / n_{s_k}$, where $K$ is the total word count. Under this scheme, the aggregate weight for all words of speaker~$i$ is $n_i \cdot (K / n_i) = K$, ensuring uniform speaker contribution. The weighted local cost becomes: $\hat{c}^{\pi}(k, t) = \omega_k \cdot c^{\pi}(k, t).$

\begin{table*}[ht]
\centering
\caption{cpWER(\%) across previous studies and four paradigms.}
\vspace{-2.5ex}
\label{tab:multi_speaker_results}
\resizebox{\textwidth}{!}{
\begin{tabular}{@{}lllcccccccccc@{}}
\toprule
\multirow{2}{*}{\textbf{Type}} & \multirow{2}{*}{\textbf{System}} & \multirow{2}{*}{\textbf{Metric}} & \multicolumn{2}{c}{\textbf{CH109}} & \multicolumn{2}{c}{\textbf{Mixer6}} & \multicolumn{2}{c}{\textbf{AMI IHM}} & \multicolumn{2}{c}{\textbf{AMI SDM}} & \multicolumn{2}{c}{\textbf{Avg}} \\ 
\cmidrule(lr){4-5} \cmidrule(lr){6-7} \cmidrule(lr){8-9} \cmidrule(lr){10-11} \cmidrule(lr){12-13}
& & & oracle & diar & oracle & diar & oracle & diar & oracle & diar & oracle & diar \\ \midrule
\multirow{5}{*}{Offline} & Kanda et al. \cite{kanda21large} & \multirow{6}{*}{\texttt{cpWER}} & --  & --  & --    & --   & --      & --  & --  & 21.2  & --  & --    \\
& Park et al. \cite{park2024enhancing} &  & --  & 26.23 & --  & --    & --  & 26.97 & --   & --    & --  & --      \\
& Cornell et al. \cite{cornell2024one}    &  & --  & --    & --  & --    & --  & --    & 21.1 & 24.5  & --  & --      \\
& VibeVoiceASR \cite{peng2026vibevoiceasr}$^\dagger$  &  & --  & 19.21 & --  & 18.46 & --  & 25.50 & --   & 42.19   & --  & 26.34   \\ 
% & DiCOW\cite{polok2026dicow}              &  & --  & --    & --  & --    & --  & --    & 17.2 & 23.6    & --  & --      \\
% & SE-DiCoW\cite{polok2026se}$^\dagger$  &  & --  & --    & --  & --    & --  & 15.3  & --   & 18.5    & --  & --      \\
& DiCOW-v3.2 \cite{polok2026dicow, polok2026se}$^\dagger$ &  & --  & 10.89 & --  & 14.43 & --  & 14.94 & --  & 20.19 & --  & 15.11 \\
& DiCOW-v3.3 \cite{polok2026dicow, polok2026se}$^\dagger$ & & 10.12 & 10.80 & 14.49 & 14.74 & 11.33 & 14.48 & 14.78 & 18.24 & 12.68 & 14.57 \\\midrule
\multirow{5}{*}{\shortstack[l]{Streaming\\(Ours)}} & Cascaded          & \multirow{5}{*}{\texttt{cpWER}}               & 38.9 & 30.72 & 39.19 & 39.97 & 49.56 & 43.54 & 53.33 & 54.83 & 45.25 & 42.27 \\
& Masked Input      &  & 23.49 & 24.27 & 34.22 & 35.39 & 26.36 & 30.66 & 36.18 & 48.75 & 30.06 & 34.77 \\
& WL-SOT (SSFv2.1~\cite{nvidia2025ssfv21} Cond.) &  & 31.17 & 32.29 & 29.65 & 28.58 & 23.8 & 31.38 & 30.02 & 41.60 & 28.66 & 33.46 \\
& SSA-v1 (SSFv2.1~\cite{nvidia2025ssfv21} Cond.) &  & 14.20 & \textbf{12.69} & 15.01 & 18.75 & 15.22 & 22.10 & 20.27 & 39.91 & 16.18 & 23.36 \\ 
& SSA-v2 (N3D~\cite{nvidia2026n3d} Cond.)   &  & 14.66 & 13.00 & 16.99 & \textbf{16.92} & 12.41 & \textbf{13.38} & 17.44 & \textbf{20.33} & 15.38 & 15.91 \\\midrule
& Streaming Diarizer \cite{medennikov25_interspeech}  & \texttt{DER} &      & 5.09  &      & 20.34 &      & 16.67 &      & 20.57 &      & 14.33 \\ \bottomrule
\end{tabular}
}
\parbox{\linewidth}{\raggedright \fontsize{8pt}{8pt}\selectfont $^\dagger$Evaluated under identical settings (audio and ground-truth) as our proposed systems; results may differ from the cited article.}
\end{table*}
%% -----------
\begin{table*}[ht]
\centering
\caption{WER (\%) on Hugging Face OpenASR leaderboard datasets~\cite{hugging_face_open_asr}.}
\vspace{-2.5ex}
\label{tab:single_speaker_results}
\resizebox{\textwidth}{!}{
\begin{tabular}{@{}l|l|c|c|cccccccc|l@{}}
\toprule
\textbf{Type} & \textbf{Model} & \textbf{Latency} & \textbf{Size} & \textbf{AMI} & \textbf{\footnotesize{Earnings}} & \textbf{\footnotesize{GigaSpeech}} & \textbf{LS-clean} & \textbf{LS-other} & \textbf{SPGI} & \textbf{Tedlium} & \textbf{Voxpopuli} & \textbf{AVG} \\
\midrule
\multirow{3}{*}{Offline} & Canary-Qwen2.5~\cite{nvidia2025canaryqwen} & $\infty$ & 2.5B & \textbf{10.19} & \textbf{10.45} & \textbf{9.43} & \textbf{1.61} & \textbf{3.1} & \textbf{1.9} & 2.71 & \textbf{5.66} & \textbf{5.63} \\
 & Whisper-Large-v3~\cite{radford2023robust} & $\infty$ & 1.55B & 15.95 & 11.29 & 10.02 & 2.01 & 3.91 & 2.94 & 3.86 & 9.54 & 7.44 \\
 & VibeVoice ASR~\cite{peng2026vibevoiceasr} & $\infty$ & 9B & 19.33 & 13.03 & 9.75 & 2.16 & 5.51 & 3.82 & \textbf{2.56} & 8.08 & 8.03 \\
\midrule
\multirow{4}{*}{Streaming} & DSM-ASR~\cite{zeghidour2025streaming} & 2.50s & 3B & 12.2 & \textbf{11.0} & \textbf{9.8} & \textbf{1.7} & \textbf{4.3} & \textbf{2.0} & \textbf{3.4} & \textbf{6.8} & \textbf{6.4} \\
 & Nemotron Speech ASR & \textbf{1.04s} & \textbf{0.6B} & \textbf{11.58} & 12.48 & 11.45 & 2.31 & 4.75 & 2.62 & 4.5 & 7.57 & 7.16 \\
 & SSA  & \textbf{1.04s} & \textbf{0.6B} & 11.62 & 14.68 & 11.49 & 2.19 & 4.76 & 2.68 & 4.65 & 7.45 & 7.44 \\
 & WL-SOT  & \textbf{1.04s} & \textbf{0.6B} & 14.64 & 37.25 & 15.51 & 9.31 & 18.72 & 15.31 & 7.55 & 17.3 & 16.95 \\
\bottomrule
\end{tabular}
}
\end{table*}

\subsubsection{Cost Calculation Mechanisms}
The DTW cost is defined via a cumulative sum:
\begin{equation}
\label{eq:dtw_recurrence}
D^{\pi}(k, t) = \hat{c}^{\pi}(k, t) + \min
\begin{cases}
    D^{\pi}(k-1, t) \\
    D^{\pi}(k-1, t-1) \\
    D^{\pi}(k, t-1) 
\end{cases}
\end{equation}
with boundary $D^{\pi}(k, 0)=\infty$ for all $k>1$.~The three allowed transitions capture distinct alignment scenarios: \textit{Vertical} $(k{-}1, t) \to (k, t)$ where multiple words align to the same frame (overlapping speech); \textit{Diagonal} $(k{-}1, t{-}1) \to (k, t)$ for one-to-one alignment; and \textit{Horizontal} $(k, t{-}1) \to (k, t)$, where one word spans multiple frames. The normalized DTW cost for permutation~$\pi$ is:
\begin{equation}
  \label{eq:dtw_total}
  \mathcal{L}_{\text{DTW}}(\pi) = \frac{D^{\pi}(K, T)}{K + T}.
\end{equation}
%% ---------------------------------------------------------------------------
For computational efficiency, the DTW cost is computed for all $P = |\Pi|$ candidate permutations simultaneously using batched computation. The outer loops over $k$ and $t$ remain sequential due to the dynamic programming (DP) data dependencies,
but all $P$ permutations are processed in a single vectorized pass.  The total
computational complexity is $\mathcal{O}(P \cdot K \cdot T)$.

%% ---------------------------------------------------------------------------
\begin{figure}[t!]
  \centering
  \includegraphics[width=0.95\linewidth]{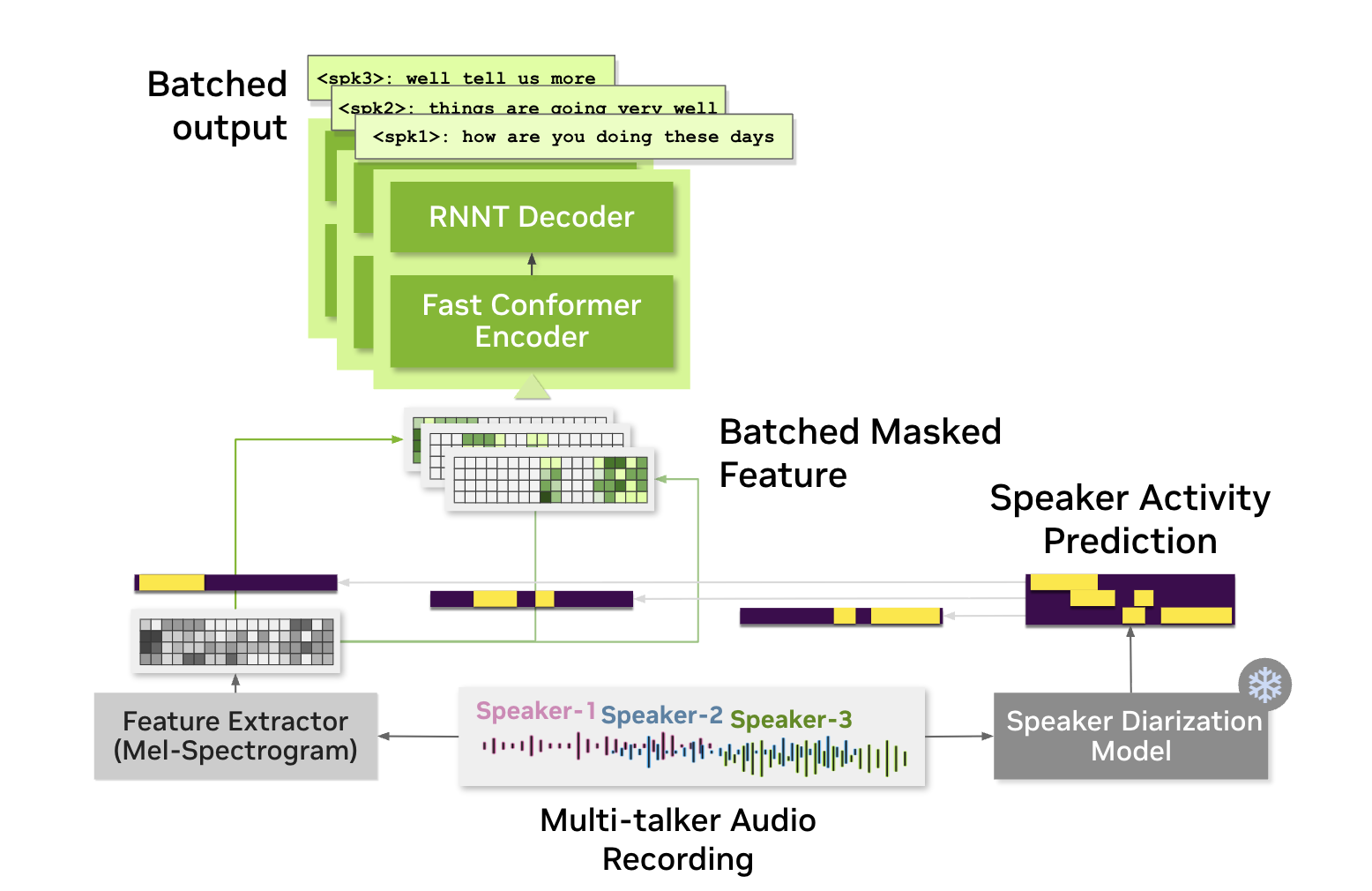}
  \vspace{-2ex}
  \caption{Diarization Conditioning Approach: Multiple instances of the ASR model are conditioned on diarization information.}
  \label{fig:diarization_conditioning}
  \vspace{-4ex}
\end{figure}
%% ---------------------------------------------------------------------------
% \subsubsection{Speaker Frequency Match Cost Calculation.}
In addition to the DTW algorithm, we also summarize each side (text vs.~RTTM) by an $N$-dimensional speaking-time vector, with one dimension per speaker. Let $\mathbf{f}^{\text{text}} \in \mathbb{R}^{N}$ denote the normalized speaking-time ratios implied by the text speakers, estimated using character counts as a proxy for duration. Similarly, let $\mathbf{f}^{\text{rttm}} \in \mathbb{R}^{N}$ denote the normalized speaking-time ratios derived from the RTTM files. Under a candidate permutation $\pi$ (mapping text speaker index $i$ to RTTM column index $\pi(i)$), we measure the mismatch by the $L_1$ distance between these two vectors after permuting the RTTM entries:
\begin{equation}
  \label{eq:freq_cost}
  \mathcal{L}_{\text{freq}}(\pi) = \sum_{i=0}^{N-1}
    \bigl|\, \mathbf{f}^{\text{text}}_i - \mathbf{f}^{\text{rttm}}_{\pi(i)} \,\bigr|.
\end{equation}
This cost is minimized when the permutation perfectly aligns the speaking-time ratios of text speakers with the RTTM columns. Finally, the optimal permutation $\pi^{*}$ is determined by minimizing the total cost, defined as the sum of the DTW and frequency costs:
\begin{equation}
  \label{eq:best_perm}
  \pi^{*} = \operatorname*{arg\,min}_{\pi \in \Pi} \Big\{ \mathcal{L}_{\text{DTW}}(\pi) + \mathcal{L}_{\text{freq}}(\pi) \Big\},
\end{equation}
where $\Pi$ is the set of all possible permutations.

\subsubsection{Model Training}
For data preparation, we adopt the data cleaning methodology described in~\cite{parksortformer} to generate short utterances containing one to four speakers. Our training corpus comprises the AMI Corpus~\cite{ami} (including Headset mix, Lapel mix, and the first channel of the Microphone array), the ICSI~\cite{icsi} dataset, the DipCo dataset~\cite{dipco}, and the Fisher English Corpus~\cite{cieri2004fisher}.

We initialize the FastConformer~\cite{rekesh2023fast} encoder using the checkpoint from~\cite{nvidia_nemotron_speech_streaming_2025}, while the RNN-T decoder is randomly initialized. Training proceeds in two stages on a single node with 8$\times$NVIDIA Tesla A100 GPUs: first, we train on short segments (8--12~s) for 50k steps; subsequently, we fine-tune for 20k steps with a learning rate of $2\times 10^{-4}$ using variable-length utterances (10--55~s) truncated from the same datasets. All other training configurations follow the specifications detailed in the SSA~\cite{wang25y_interspeech} paper.

\subsection{Diarization-Conditioned Multi-Instance Schemes}
The diarization-conditioned architecture~\cite{polok2026dicow, polok2026se} employs a parallel, multi-instance configuration where ASR models are explicitly conditioned on diarization embeddings or speaker representations. In this work, we adopt Self-Speaker Adaptation (SSA) ASR~\cite{wang25y_interspeech}, a streaming-capable implementation built upon the pretrained Nemotron Speech streaming model~\cite{nvidia_nemotron_speech_streaming_2025}. We categorize SSA as a diarization conditioning approach because it directly ingests diarization predictions to spawn multiple ASR instances, each dedicated to transcribing a specific speaker.

\begin{itemize}
    \item \textbf{Pros:}~Delivers superior performance on overlapping speech; maintains negligible regression in single-speaker ASR accuracy due to the speaker-specific decoding strategy.
    \item \textbf{Cons:}~Requires specialized fine-tuning to integrate speaker conditioning mechanisms; exhibits a larger memory footprint due to the multi-instance approach; incurs high computational costs when transcribing sessions with a large number of speakers.
\end{itemize}
\subsubsection{Model Training}
We closely follow the training methodology presented in~\cite{wang25y_interspeech}, though our model also incorporates a single-speaker dataset from Granary~\cite{raokoluguri25_interspeech}. In addition to the training datasets used in~\cite{wang25y_interspeech} (Fisher English Corpus~\cite{cieri2004fisher} and LibriSpeechMix~\cite{kanda2020serialized, kahn2020librilight}), we augment our training data with the AMI Corpus~\cite{ami}, ICSI~\cite{icsi}, and NOTSOFAR-1~\cite{vinnikov24_interspeech} datasets. All training configurations---including tokenizers, optimizers, learning rate schedulers, and hardware specifications---are identical to those detailed in the SSA~\cite{wang25y_interspeech} study.

\section{Experiments and Results}

To evaluate the impact of these architectural choices, we assess the four paradigms on several conversational datasets: CH109, a two-speaker subset (109 files) of the Call-Home American English Dataset~\cite{canavan1997CALLHOME}; Mixer6~\cite{Mixer6}; and the AMI Meeting Corpus~\cite{ami} under both Individual Headset Mix~(IHM) and Single Distant Microphone~(SDM) conditions.

Table~\ref{tab:multi_speaker_results} presents the concatenated minimum permutation word error rate (cpWER)~\cite{watanabe2020chime} across these four datasets. We include results from recently published multi-speaker ASR systems to benchmark our proposed methods. For open-source models such as VibeVoiceASR~\cite{peng2026vibevoiceasr} and the latest DiCoW-v3.2~\cite{polok2026dicow, polok2026se}, we perform evaluations using identical audio files and annotations to ensure a fair comparison. The \textit{oracle} columns report the performance of our proposed systems using ground-truth diarization, effectively removing degradation caused by diarization errors. Consistent with our earlier observations, the SSA approach (diarization conditioning) achieves the best performance. The WL-SOT model slightly outperforms the masked input approach, while the cascaded method yields the lowest accuracy for multi-speaker ASR. Table~\ref{tab:single_speaker_results} reports the WER on the \textit{Hugging Face OpenASR leaderboard}~\cite{srivastav2025openasrleaderboardreproducible} datasets. Notably, while the WL-SOT model exhibits significant degradation in single-speaker accuracy, SSA largely preserves the performance of the Nemotron Speech ASR~\cite{nvidia_nemotron_speech_streaming_2025} base model.

\section{Conclusion}
In this work, we formalize the architectural trade-offs in streaming multi-talker ASR and introduce the PI-DTW algorithm to resolve label alignment for WL-SOT training. While specialized diarization conditioning currently holds the accuracy advantage, resolving SOT's alignment bottleneck enables future exploration into true end-to-end joint optimization models. Future work will focus on scaling the WL-SOT training data and supporting speaker-tagging natively in Attention-based Encoder-Decoder models.

\section{Generative AI Use Disclosure}
In the preparation of this manuscript, the authors used generative AI services solely for technical assistance, including refining English grammar, correcting LaTeX syntax, and formatting data tables. All substantive content, original research, and interpretations were produced by the authors. The authors have reviewed and edited all AI-assisted outputs and take full responsibility for the accuracy and integrity of the final work.

\bibliographystyle{IEEEtran}
\bibliography{reference}

% \end{document}

\end{document}